\documentclass[
12pt,
preprint,preprintnumbers,nofootinbib,
groupedaddress,superscriptaddress,amsmath,amssymb]{revtex4}
\usepackage{graphicx}
\usepackage{dcolumn}
\usepackage{bm}
\usepackage{amssymb}
\usepackage{amsmath}
\usepackage{epsfig}    
\usepackage{color}
\usepackage{hhline}

\def\be{\begin{equation}}
\def\ee{\end{equation}}
\newcommand{\bea}{\begin{eqnarray}}
\newcommand{\eea}{\end{eqnarray}}
\newcommand{\nn}{\nonumber}

\numberwithin{equation}{section}

\begin{document}
{\begin{flushright}{CTP-SCU/2022001,  APCTP Pre2022 - 001}\end{flushright}}

\title{A radiative seesaw model in a supersymmetric modular $A_4$ group}

\author{Takaaki Nomura}
\email{nomura@scu.edu.cn}
\affiliation{College of Physics, Sichuan University, Chengdu 610065, China}

\author{Hiroshi Okada}
\email{hiroshi.okada@apctp.org}
\affiliation{Asia Pacific Center for Theoretical Physics (APCTP) - Headquarters San 31, Hyoja-dong,
Nam-gu, Pohang 790-784, Korea}

\date{\today}

\begin{abstract}
We propose a supersymmetric radiative seesaw model with modular $A_4$ symmetry.
Thanks to contributions of supersymmetric partners to one-loop diagrams generating neutrino masses, we successfully fit neutrino data and obtain predictions in case of normal hierarchy in a minimal framework that would not be realized in a non-supersymmetric model.
We show a several predictive figures and demonstrate a best fit benchmark point through $\chi^2$ analysis. 
\end{abstract}
\maketitle
\newpage

\section{Introduction}
{The neutrino sector is described by several observables such as
neutrino masses containing two or three non-zero mass eigenvalues;  three mixing angles inducing neutrino oscillations; CP phases including Dirac CP and Majorana phases that are not yet perfectly confirmed by experiments. 
We can come up with various ideas in order not only to reproduce experimental data but also to predict unknown observables such as phases.
The neutrino mass model is sometimes constructed under a radiatively induced mass scenario~\cite{Ma:2006km}, called radiative seesaw,
in which we can induce the neutrino mass at loop levels while tree level mass is forbidden.
Such a model would be natural in order to explain the tiny neutrino masses, retaining not-so-small Yukawa couplings.
It also sometimes accommodates a dark matter (DM) candidate and potentially causes lepton flavor violation that could leads us to more intriguing phenomenology. 
This kind of model usually requires additional symmetry to guarantee the radiative seesaw mechanism and stability of DM.  
In order to get predictions in the lepton sector, non-Abelian discrete flavor symmetries frequently plays an outstanding role.
The symmetries (that has currently been developed with the help of modular symmetry) would perfectly match with the radiative seesaw model, because a remnant or part of symmetry plays a role in replacing the additional symmetry  to guarantee the radiative seesaw mechanism and stability of DM.  }
 

The modular flavor symmetry has been proposed by 2017 in ref.~\cite{Feruglio:2017spp, deAdelhartToorop:2011re}. 
After that, a lot of ideas have been come up with in order to realize predictive models.
For example, 
the modular $A_4$ flavor symmetry has been discussed in refs.~\cite{Feruglio:2017spp, Criado:2018thu, Kobayashi:2018scp, Okada:2018yrn, Nomura:2019jxj, Okada:2019uoy, deAnda:2018ecu, Novichkov:2018yse, Nomura:2019yft, Okada:2019mjf,Ding:2019zxk, Nomura:2019lnr,Kobayashi:2019xvz,Asaka:2019vev,Zhang:2019ngf, Gui-JunDing:2019wap,Kobayashi:2019gtp,Nomura:2019xsb, Wang:2019xbo,Okada:2020dmb,Okada:2020rjb, Behera:2020lpd, Behera:2020sfe, Nomura:2020opk, Nomura:2020cog, Asaka:2020tmo, Okada:2020ukr, Nagao:2020snm, Okada:2020brs, Yao:2020qyy, Chen:2021zty, Kashav:2021zir, Okada:2021qdf, deMedeirosVarzielas:2021pug, Nomura:2021yjb, Hutauruk:2020xtk, Ding:2021eva, Nagao:2021rio, king, Okada:2021aoi, Nomura:2021pld, Kobayashi:2021pav, Dasgupta:2021ggp, Liu:2021gwa}, 
$S_3$  in refs.~\cite{Kobayashi:2018vbk, Kobayashi:2018wkl, Kobayashi:2019rzp, Okada:2019xqk, Mishra:2020gxg, Du:2020ylx}, 
$S_4$  in refs.~\cite{Penedo:2018nmg, Novichkov:2018ovf, Kobayashi:2019mna, King:2019vhv, Okada:2019lzv, Criado:2019tzk,
Wang:2019ovr, Zhao:2021jxg, King:2021fhl, Ding:2021zbg, Zhang:2021olk, gui-jun, Nomura:2021ewm}, 
$A_5$ in refs.~\cite{Novichkov:2018nkm, Ding:2019xna,Criado:2019tzk},
double covering of $A_4$  in refs.~\cite{Liu:2019khw, Chen:2020udk, Li:2021buv}, 
double covering of $S_4$  in refs.~\cite{Novichkov:2020eep, Liu:2020akv},   and
double covering of $A_5$  in refs.~\cite{Wang:2020lxk, Yao:2020zml, Wang:2021mkw, Behera:2021eut}.
Other types of modular symmetries have also been proposed to understand masses, mixings, and phases of the standard model (SM) in refs.~\cite{deMedeirosVarzielas:2019cyj, Kobayashi:2018bff,Kikuchi:2020nxn, Almumin:2021fbk, Ding:2021iqp, Feruglio:2021dte, Kikuchi:2021ogn, Novichkov:2021evw, Kikuchi:2021yog, Novichkov:2022wvg}.~\footnote{Here, we provide useful review references for beginners~\cite{Altarelli:2010gt, Ishimori:2010au, Ishimori:2012zz, Hernandez:2012ra, King:2013eh, King:2014nza, King:2017guk, Petcov:2017ggy}.}
Different applications to physics such as dark matter and origin of CP violation are found in refs.~\cite{Kobayashi:2021ajl, Nomura:2019jxj, Nomura:2019yft, Nomura:2019lnr, Okada:2019lzv, Baur:2019iai, Kobayashi:2019uyt, Novichkov:2019sqv,Baur:2019kwi, Kobayashi:2020hoc, Tanimoto:2021ehw}.
Mathematical study such as possible correction from K\"ahler potential, systematic analysis of the fixed points,
moduli stabilization are discussed in refs.~\cite{Chen:2019ewa, deMedeirosVarzielas:2020kji, Ishiguro:2020tmo, Abe:2020vmv}.
It is recently studied that a scenario to derive four-dimensional modular flavor symmetric models from higher dimensional theory by assuming the compactification consistent with the modular symmetry~\cite{Kikuchi:2022txy}.

In this letter, we extend our original radiative seesaw model analyzed under the non-supersymmetric  (non-SUSY) framework~\cite{Nomura:2019jxj}
to the one under the SUSY version considering contributions from superpartner particles. Thanks to contributions from supersymmetric partners to diagrams generating neutrino masses, we successfully fit neutrino data and obtain predictions  in case of normal hierarchy in a minimal framework that would not be realized in a non-supersymmetric model.
We show several predictive figures and demonstrate a best fit benchmark point through $\chi^2$ analysis. 

This letter is organized as follows.
In Sec.~\ref{sec:realization},   we explain our model setup under the modular $A_4$ symmetry, formulating valid mass matrix and their mixings and the neutrino sector. In Sec~\ref{sec:NA1}, we carry out our  $\chi^2$ numerical analysis scanning free parameters.
Finally we conclude and discuss in Sec.~\ref{sec:conclusion}.

\begin{center} 
\begin{table}[tb]
\begin{tabular}{|c||c|c|c|c|c|c|c||c|c|c|c|c|c||}\hline\hline  
  & ~$L_e$~& ~$L_{\mu}$~ & ~$L_{\tau}$~& ~$e^c$~& ~$\mu^c$~ & ~$\tau^c$~ & ~$N^c$~  & ~$H_1$~& ~$H_2$~& ~$\eta_1$~& ~$\eta_2$~ &~$\chi$~  
  \\\hline 
 $SU(2)_L$ & $\bm{2}$  & $\bm{2}$  & $\bm{2}$ & $\bm{1}$   & $\bm{1}$  & $\bm{1}$ & $\bm{1}$ & $\bm{2}$   & $\bm{2}$ & $\bm{2}$  & $\bm{2}$ & $\bm{1}$   \\\hline 
$U(1)_Y$ & $-\frac12$ & $-\frac12$ & $-\frac12$  & $1$& $1$ & $1$  & $0$  & $\frac12$   & $-\frac12$  & $\frac12$   & $-\frac12$ & $0$    \\\hline
 $A_4$ & $1$ & $1'$ & $1''$ & $1$ & $1'$ & $1''$ & $3$ & $1$ & $1$ & $1$   & $1$& $1$    \\\hline
 $k$ & $0$ & $0$ & $0$ & $0$ & $0$ & $0$ & $-1$ & $0$ & $0$ & $-1$  & $-3$ & $-3$    \\\hline
\end{tabular}
\caption{Superfield contents
and their charge assignments under $SU(2)_L\times U(1)_Y\times A_{4}$, where $-k$ is the number of modular weight.}
\label{tab:fields}
\end{table}
\end{center}

\section{Model} 
\label{sec:realization}
In this section we review our model with modular $A_4$ symmetry in a framework of SUSY.
We construct the model as minimal assignment as possible.
Thus we give zero modular weight to leptons $L_{(e,\mu,\tau)}, \ (e^c,\mu^c,\tau^c)$ and two Higgs doublets $H_1$ and $H_2$. 
When we respectively assign $A_4$ singlet representations $(1,1',1'')$ and  $(1,1'',1')$ to $L_{(e,\mu,\tau)}$ and  $(e^c,\mu^c,\tau^c )$ , the charged-lepton mass matrix is diagonal. Therefore, the observed lepton mixing arises from the neutrino sector only. %
{To induce neutrino masses at one-loop level minimally, we introduce the SM singlet superfields $N^c$ and $\chi$, and inert doublet superfields $\eta_{1,2}$. Here,  $N^c$ is assigned to triplet under $A_4$ with $-1$ modular weight and corresponds to heavy Majorana neutrino. }%
The second inert doublet $\eta_2$ plays the same role in generating the mass terms for $\eta_{1,2}$ as the MSSM Higgs sector $H_{1,2}$. The singlet $\chi$ is required in order to connect the boson loop in neutrino mass generating diagram.~\footnote{SUSY does not allow to induce the interaction $(H_1^\dag\eta_1)^2$ in a renormalizable theory.}
Superfields $H_{1,2}$, $\eta_{1,2}$ and $\chi$  are chosen to be trivial singlet under $A_4$, while zero modular weight for $H_{1,2}$ and $(-1,-3,-3)$ weights for $(\eta_1,\eta_2,\chi)$ are assigned;  we choose the assignment to make the model minimal. 
All the charge assignment and field contents are summarized in Table~\ref{tab:fields}.
Under these symmetries, our valid superpotential is given as follows (formulas in modular $A_4$ framework are summarized in the Appendix):
\begin{align}
{\cal W} &=
\sum_{\ell=e,\mu,\tau} y_{\ell} \ell^c L_{\ell} H_1\nn\\
&+a_\eta (y_1 N^c_1 +y_2 N^c_3+y_3 N^c_2) L_e \eta_1
+b_\eta (y_2 N^c_2 +y_1 N^c_3+y_3 N^c_1) L_\mu \eta_1
+c_\eta (y_3 N^c_3 +y_1 N^c_2+y_2 N^c_1) L_\tau \eta_1\nn\\
&+ m_N \left[
y_1(2 N^c_{1} N^c_{1} -N^c_{2} N^c_{3}-N^c_{3} N^c_{2})
+
y_2(2 N^c_{2} N^c_{2} -N^c_{1} N^c_{3}-N^c_{3} N^c_{1})
+
y_3(2 N^c_{3} N^c_{3} -N^c_{1} N^c_{2}-N^c_{2} N^c_{1})\right]\nn\\
&+\mu_1 Y^{(6)}_1 H_1 \eta_2 \chi + \mu_2 Y^{(4)}_1 H_2 \eta_1 \chi
+\mu_\chi Y^{(6)}_1 \chi\chi
+\mu_H  H_1H_2
+\mu_{\eta} Y^{(4)}_1 \eta_1 \eta_2, \label{eq:w-lep}
\end{align}
where $Y^{(2)}_3\equiv (y_1,y_2,y_3)^T$, and we imposed R-parity forbidding the R-parity violating terms such as $L_e H_1,\ L_e L_\mu\tau^c$.
Note here that oddness of modular weight provides oddness under accidental $Z_2$ symmetry since all modular forms have even modular weight in modular $A_4$ framework 
and sum of modular weight of superfields should be even to make modular weight of a term zero.

Then, the valid renormalizable soft SUSY breaking Lagrangian is found as follows:
\begin{align}
-{\cal L}_{\rm soft} 
&=
A_1 Y^{(6)}_1 H_1 \eta_2 \chi + A_2 Y^{(4)}_1 H_2 \eta_1 \chi 
+ m^2_{H_1} |H_1|^2+ m^2_{H_2} |H_2|^2
\nn\\
&+ B_{\tilde N}^2 \Bigl[
y_1(2 \tilde N^c_{1}\tilde  N^c_{1} -\tilde  N^c_{2}\tilde  N^c_{3}-\tilde N^c_{3}\tilde  N^c_{2})
+
y_2(2 \tilde  N^c_{2} \tilde N^c_{2} -\tilde N^c_{1} \tilde N^c_{3}-\tilde N^c_{3} \tilde N^c_{1}) \nonumber \\
& \qquad \quad +
y_3(2 \tilde N^c_{3} \tilde N^c_{3} -\tilde N^c_{1} \tilde N^c_{2}-\tilde N^c_{2} \tilde N^c_{1}) \Bigr]
\nn\\
&+B_H^2 H_1 H_2 + B_\eta^2 Y^{(4)}_1 \eta_1 \eta_2+B_2^2 Y^{(6)}_1 \chi \chi, \label{eq:lag-lep}
\end{align}
where all fields are supposed to be scalars; $\tilde N^c$'s are gauge singlet sneutrinos.
{Note that we also have SUSY breaking terms associated with sleptons, squarks and gauginos in the MSSM. 
In our analysis, however, we do not discuss super partner of  the SM fermions and gauge bosons assuming they are heavy enough to avoid experimental constraints, and focus on 
neutrino mass generating mechanism.}

\subsection{$N^c$ mass matrix}
The mass matrix of right-handed neutral fermions is straightforwardly found from the suerpotential as follows:
\begin{align}
{\cal M_N} &=
m_N
\left[\begin{array}{ccc}
2 y_1 & -y_3 & -y_2 \\ 
-y_3 &2y_2  & -y_1   \\ 
-y_2 & -y_1 & 2y_3  \\ 
\end{array}\right]
\equiv m_N {\cal Y}
,\label{eq:mn}
\end{align}
where  ${\cal M_N}$ is diagonalized by $D_N \equiv U_N{\cal M_N} U_N^T\equiv (M_{1},M_{2},M_{3})$, where  $U_N$ is a unitary matrix, and we define the mass eigen-field $\psi^c$ as $N^c\equiv U^T_N \psi^c_N$.

\subsection{$\tilde N^c$ mass matrix}
The mass matrix of $\tilde N^c$ arises from the $F$ and soft SUSY breaking terms;
\begin{align}
-{\cal L}_{\tilde N}=|m_N|^2(\tilde N^c)^*{\cal Y}^*{\cal Y} \tilde N^c +B_{\tilde N}^2 \tilde N^c {\cal Y} \tilde N^c +{\rm h.c.}.
 \end{align}
Here, we write $\tilde N^c_i\equiv (\phi_R+i\phi_I)_i/\sqrt2$.
Then, mass terms are given by
\begin{align}
-{\cal L}_{\tilde N}&=
\phi_{R_i}^T \left(\frac{|m_N|^2}2 {\cal Y}^*{\cal Y} + B_{\tilde N}^2{\cal Y}+{\rm h.c.} \right)_{ij} \phi_{R_j}
+
\phi_{I_i}^T \left(\frac{|m_N|^2}2 {\cal Y}^*{\cal Y} - B_{\tilde N}^2{\cal Y}+{\rm h.c.} \right)_{ij} \phi_{I_j}\nn\\
&\equiv 
\phi_{R_i}^T M_{R_{ij}}^2 \phi_{R_j}
+
\phi_{I_i}^T M_{I_{ij}}^2 \phi_{I_j}.
\end{align}
The mass matrix $M_{R(I)}^2$ is diagonalized such that $D_{R(I)}^2 \equiv O_{R(I)} M_{R(I)}^2 O_{R(I)}^T\equiv (m^2_{R(I)_1},m^2_{R(I)_2},m^2_{R(I)_3})$, where $O_{R(I)}$ is an orthogonal matrix.
Furthermore, we define the mass eigen-fields $\tilde n_{R,I}$ then we have $\phi_{R(I)} = O^T_{R(I)} \tilde n_{R(I)}$.

\subsection{$\chi-\eta$ mixing}
Scalar fields $\chi$ and $\eta_{1,2}$ mix each other due to the soft-breaking terms associated with $A_1$ and $A_2$  after the EW symmetry breaking by non-zero VEV of $H_{1,2}$.
Here, we assume the mixing between $\chi$ and $\eta_1$ is only active for simplicity. Note, however, that this assumption does not affect the mechanism of neutrino mass matrix. \\
{\it Scalar boson mixing} is then defined by 
\begin{align}
& \chi_{R}=c_{R} H_1 - s_R H_2,\quad 
\eta_{R_1} =s_R H_1 + c_R H_2,\nn\\
& \chi_{I}=c_{I} A_1 - s_I A_2,\quad 
\eta_{I_1} =s_I A_1 + c_I A_2,
\end{align}
where $s_{R/I}(c_{R/I})$ is the short-hand symbol of $\sin\theta_{R/I}(\cos\theta_{R/I})$, and $H_{1,2}$ and $A_{1,2}$ are respectively CP-even and odd mass eigenstates.\\

We also consider the mixing between the fermionic superpartners of $\chi$ and $\eta$ denoted by $\tilde \chi$ and $\tilde \eta_1$.  
{\it Fermion mixing} is defined by 
\begin{align}
& \tilde\chi =c_\alpha \psi_\chi - s_\alpha \psi_\eta,
\quad 
\tilde\eta_1 = s_\alpha \psi_\chi + c_\alpha \psi_\eta,
\end{align}
where $s_\alpha(c_\alpha)$ is the short-hand symbol of $\sin\theta_{\alpha}(\cos\theta_{\alpha})$, and $\psi_\chi$ and $\psi_\eta$ are  mass eigenstates.

\begin{figure}[tb]\begin{center}
\includegraphics[width=65mm]{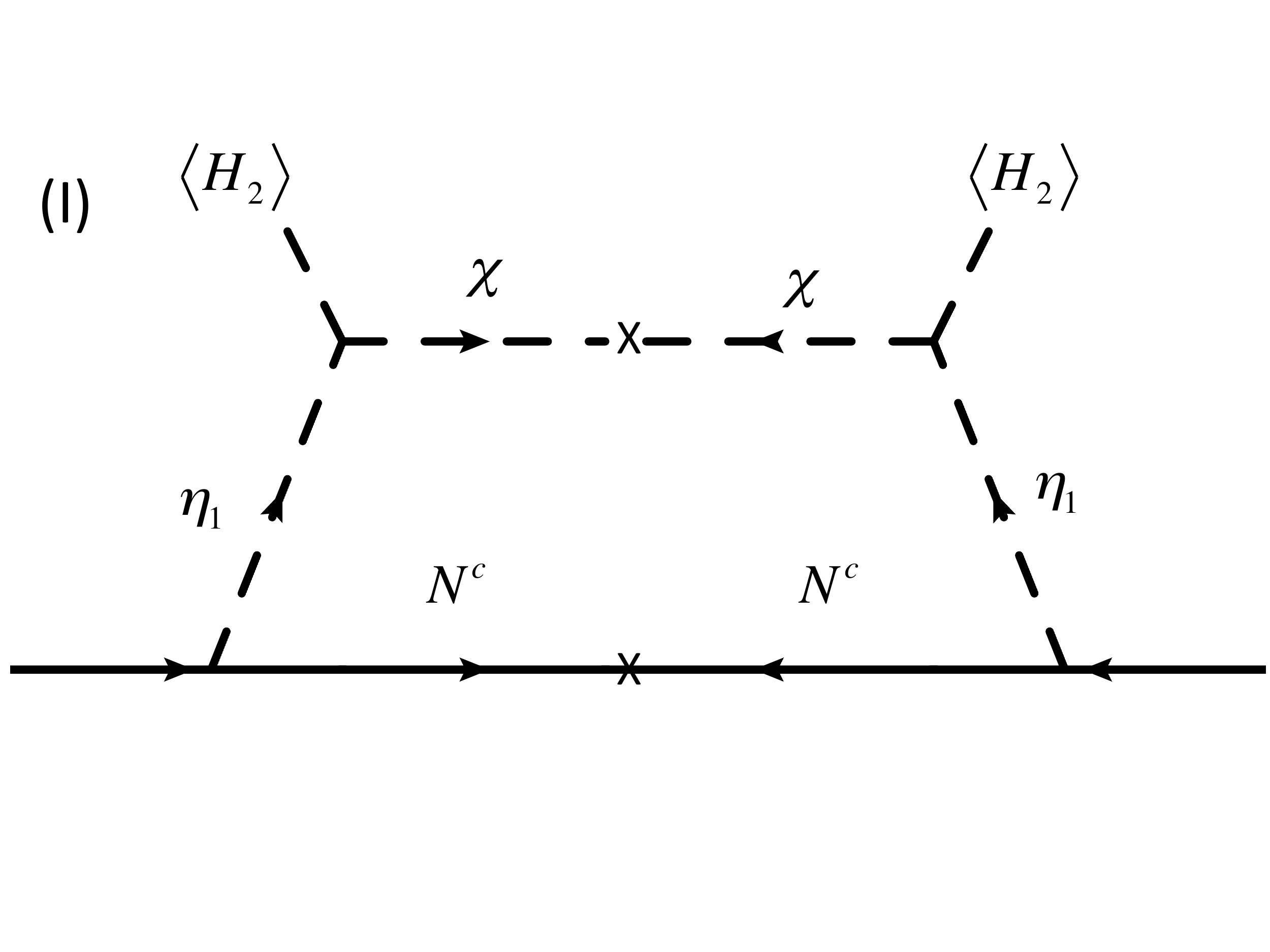} \quad
\includegraphics[width=65mm]{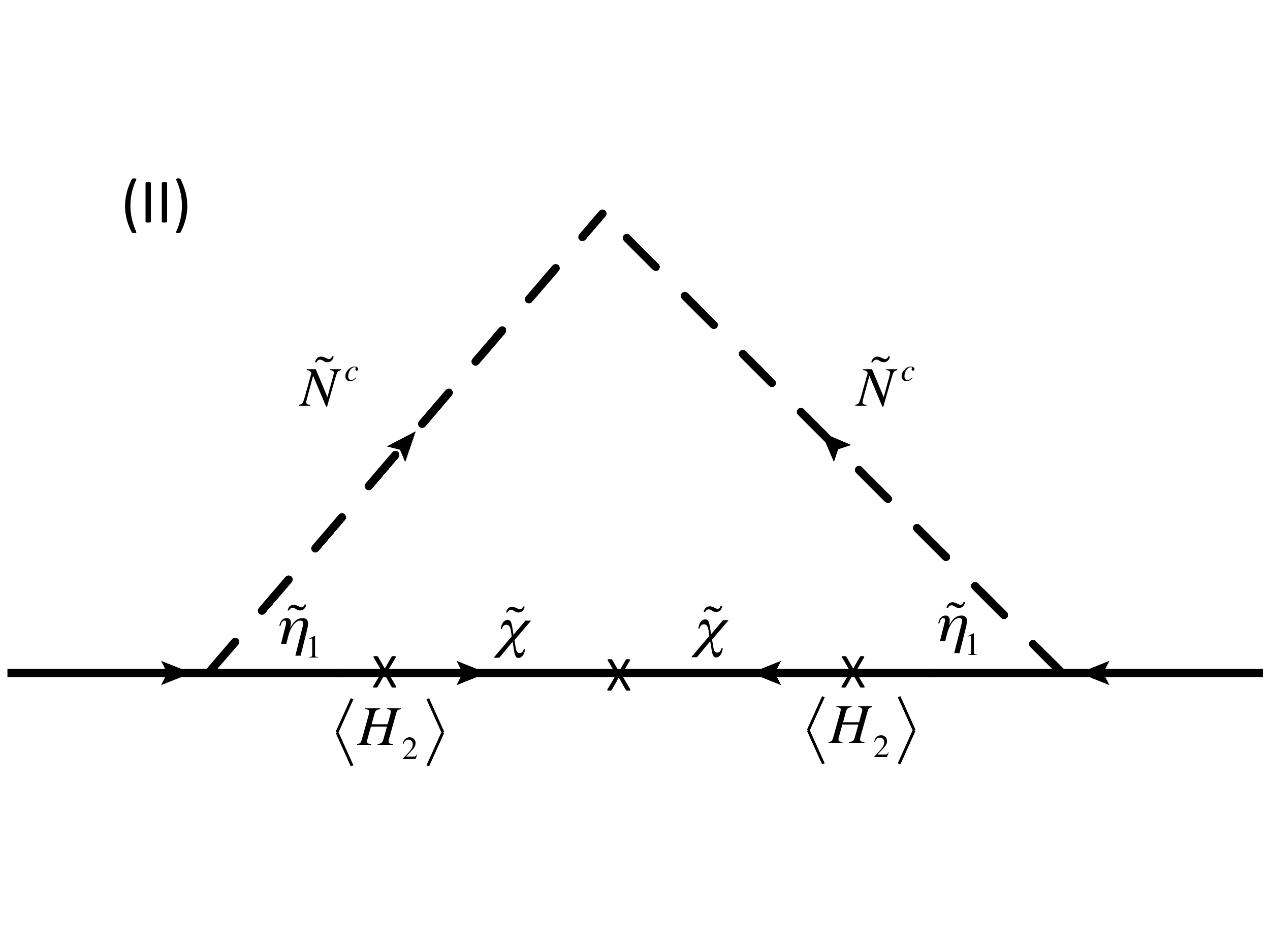}
\caption{One loop diagrams generating neutrino mass where we write particles inside loops by flavor eigenstates.}   
\label{fig:diagram}\end{center}\end{figure}

\subsection{Neutrino sector}
Now we can discuss the neutrino sector estimating neutrino mass at one-loop level.
Our valid renormalizable Lagrangian is explicitly given by
\begin{align}
-{\cal L}^\nu &=
\frac{1}{\sqrt2} (\psi^c_N)^T_i (U_N)_{ia} Y_{ab} \nu_b (s_R H_1 +c_R H_2)
+\frac{i}{\sqrt2} (\psi^c_N)^T_i (U_N)_{ia} Y_{ab} \nu_b (s_I A_1 +c_I A_2)\nn\\
&+
\frac{1}{\sqrt2} (\tilde n_R)^T_i (O_R)_{ia} Y_{ab} \nu_b (s_\alpha \psi_\chi +c_\alpha \psi_\eta)
+\frac{i}{\sqrt2} (\tilde n_I)^T_i (O_I)_{ia} Y_{ab} \nu_b (s_\alpha \psi_\chi +c_\alpha \psi_\eta) +{\rm h.c.}.\label{eq:neut}
\end{align}
{These interactions induce neutrino mass at one-loop level via the diagrams in Fig.~\ref{fig:diagram}}.
Calculating the diagrams the neutrino mass matrix is given by $m_\nu\equiv m_\nu^{(I)} + m_\nu^{(II)}$;
\begin{align}
(m_\nu^{(I)})_{ab}
&=
-\frac{1}{2(4\pi)^2} Y^T_{a\beta} (U^T_N)_{\beta i} D_{N_i} (U_N)_{i\alpha} Y_{\alpha b}\nn\\
&\times
\left[
s^2_R F(m_{H_1},M_i) - s^2_I F(m_{A_1},M_i)
+
c^2_R F(m_{H_2},M_i) - c^2_I F(m_{A_2},M_i)
\right],\label{eq:neut1}\\
(m_\nu^{(II)})_{ab}
&=
-\frac{1}{2(4\pi)^2} Y^T_{a\beta} Y_{\alpha b}\nn\\
&\times
\left[
M_{\psi_\chi} s^2_\alpha[ (O_R)_{\beta i} F(m_{R_i},M_{\psi_\chi}) (O_R)^T_{i\alpha} 
 -  (O_I)_{\beta i} F(m_{I_i},M_{\psi_\chi}) (O_I)^T_{i\alpha} ]\right.\nn\\
&\left. 
+
M_{\psi_\eta} c^2_\alpha[ (O_R)_{\beta i} F(m_{R_i},M_{\psi_\eta}) (O_R)^T_{i\alpha} 
 -  (O_I)_{\beta i} F(m_{I_i},M_{\psi_\eta}) (O_I)^T_{i\alpha} ]
 \right],\label{eq:neut2}\\
 F(m_a,m_b)
 &=
 \int_0^x dx\ln\left[ x \left(\frac{m_a^2}{m_b^2}-1\right) + 1\right],
\end{align}
where $(m_\nu^{(I)})_{ab}$ and $(m_\nu^{(II)})_{ab}$ denote contribution from diagram (I) and (II) respectively.
Here, $(m_\nu^{(I)})_{ab}$ comes from the first line of  Eq.(\ref{eq:neut}), while
$(m_\nu^{(II)})_{ab}$ comes from the second line of  Eq.(\ref{eq:neut}).
Then, $m_\nu$ is diagonalized by a unitary matrix $U_{\rm PMNS}$; $U_{\rm PMNS}^T m_\nu U_{\rm PMNS}\equiv {\rm diag}(m_1,m_2,m_3)$.  

Several experimental data are given as follows. 
We write mass square difference 
\begin{align}
(\mathrm{NH}):\  \Delta m_{\rm atm}^2 =m_3^2 - m_1^2,
\quad
(\mathrm{IH}):\    \Delta m_{\rm atm}^2 =m_2^2 - m_3^2,
 \end{align}
where $\Delta m_{\rm atm}^2$ is atmospheric neutrino mass square difference, and NH and IH represent the normal hierarchy and the inverted hierarchy, respectively. 
Solar mass square difference is given as follows:
\begin{align}
\Delta m_{\rm sol}^2=m_2^2 - m_1^2,
 \end{align}
 which can be compared to the observed value.
 %
$U_{\rm PMNS}$ is parametrized by three mixing angle $\theta_{ij} (i,j=1,2,3; i < j)$, one CP violating Dirac phase $\delta_{CP}$,
and two Majorana phases $\{\alpha_{21}, \alpha_{32}\}$ as follows:
\begin{equation}
U_{\rm PMNS}= 
\begin{pmatrix} c_{12} c_{13} & s_{12} c_{13} & s_{13} e^{-i \delta_{CP}} \\ 
-s_{12} c_{23} - c_{12} s_{23} s_{13} e^{i \delta_{CP}} & c_{12} c_{23} - s_{12} s_{23} s_{13} e^{i \delta_{CP}} & s_{23} c_{13} \\
s_{12} s_{23} - c_{12} c_{23} s_{13} e^{i \delta_{CP}} & -c_{12} s_{23} - s_{12} c_{23} s_{13} e^{i \delta_{CP}} & c_{23} c_{13} 
\end{pmatrix}
\begin{pmatrix} 1 & 0 & 0 \\ 0 & e^{i \frac{\alpha_{21}}{2}} & 0 \\ 0 & 0 & e^{i \frac{\alpha_{31}}{2}} \end{pmatrix},
\end{equation}
where $c_{ij}$ and $s_{ij}$ stands for $\cos \theta_{ij}$ and $\sin \theta_{ij}$ respectively. 
Then, each of mixing is given in terms of the component of $U_{\mathrm{PMNS}}$ as follows:
\begin{align}
\sin^2\theta_{13}=|(U_{\mathrm{PMNS}})_{13}|^2,\quad 
\sin^2\theta_{23}=\frac{|(U_{\mathrm{PMNS}})_{23}|^2}{1-|(U_{\mathrm{PMNS}})_{13}|^2},\quad 
\sin^2\theta_{12}=\frac{|(U_{\mathrm{PMNS}})_{12}|^2}{1-|(U_{\mathrm{PMNS}})_{13}|^2}.
\end{align}
Also, we compute the Jarlskog invariant, $\delta_{CP}$ derived from PMNS matrix elements $U_{\alpha i}$:
\begin{equation}
J_{CP} = \text{Im} [U_{e1} U_{\mu 2} U_{e 2}^* U_{\mu 1}^*] = s_{23} c_{23} s_{12} c_{12} s_{13} c^2_{13} \sin \delta_{CP},
\end{equation}
and the Majorana phases are also estimated in terms of other invariants $I_1$ and $I_2$:
\begin{equation}
I_1 = \text{Im}[U^*_{e1} U_{e2}] = c_{12} s_{12} c_{13}^2 \sin \left( \frac{\alpha_{21}}{2} \right), \
I_2 = \text{Im}[U^*_{e1} U_{e3}] = c_{12} s_{13} c_{13} \sin \left( \frac{\alpha_{31}}{2} - \delta_{CP} \right).
\end{equation}
In addition, the effective mass for the neutrinoless double beta decay is given by
\begin{align}
\langle m_{ee}\rangle=  |m_1 \cos^2\theta_{12} \cos^2\theta_{13}+m_2 \sin^2\theta_{12} \cos^2\theta_{13}e^{i\alpha_{21}}+m_3 \sin^2\theta_{13}e^{i(\alpha_{31}-2\delta_{CP})}|,
\end{align}
where its observed value could be measured by KamLAND-Zen in future~\cite{KamLAND-Zen:2016pfg}. 
We will adopt the neutrino experimental data in NuFit5.0~\cite{Esteban:2018azc} in oder to perform the numerical $\chi^2$ analysis. 

\section{Numerical analysis \label{sec:NA1}}
In this section, we show our numerical $\chi^2$ analysis where we set the following ranges for free parameters,
\begin{align}
&\{ a_\eta, b_\eta, c_\eta, s_R,s_I,s_\alpha \}   \in [10^{-3},1],\\
& \{m_N, B_{\tilde N}, m_{H_1}, m_{H_2}, m_{A_1}, m_{A_2}, M_{\psi_\chi}, M_{\psi_\eta} \} \in [10^2,10^5]{\rm GeV}, 
\end{align}
where modulus $\tau$ runs over the fundamental region. 
Notice here that we focus on the case of NH because we find it difficult to obtain the allowed region within $\sqrt{\chi^2}=6$ in the IH case.

\begin{figure}[tb]\begin{center}
\includegraphics[width=75mm]{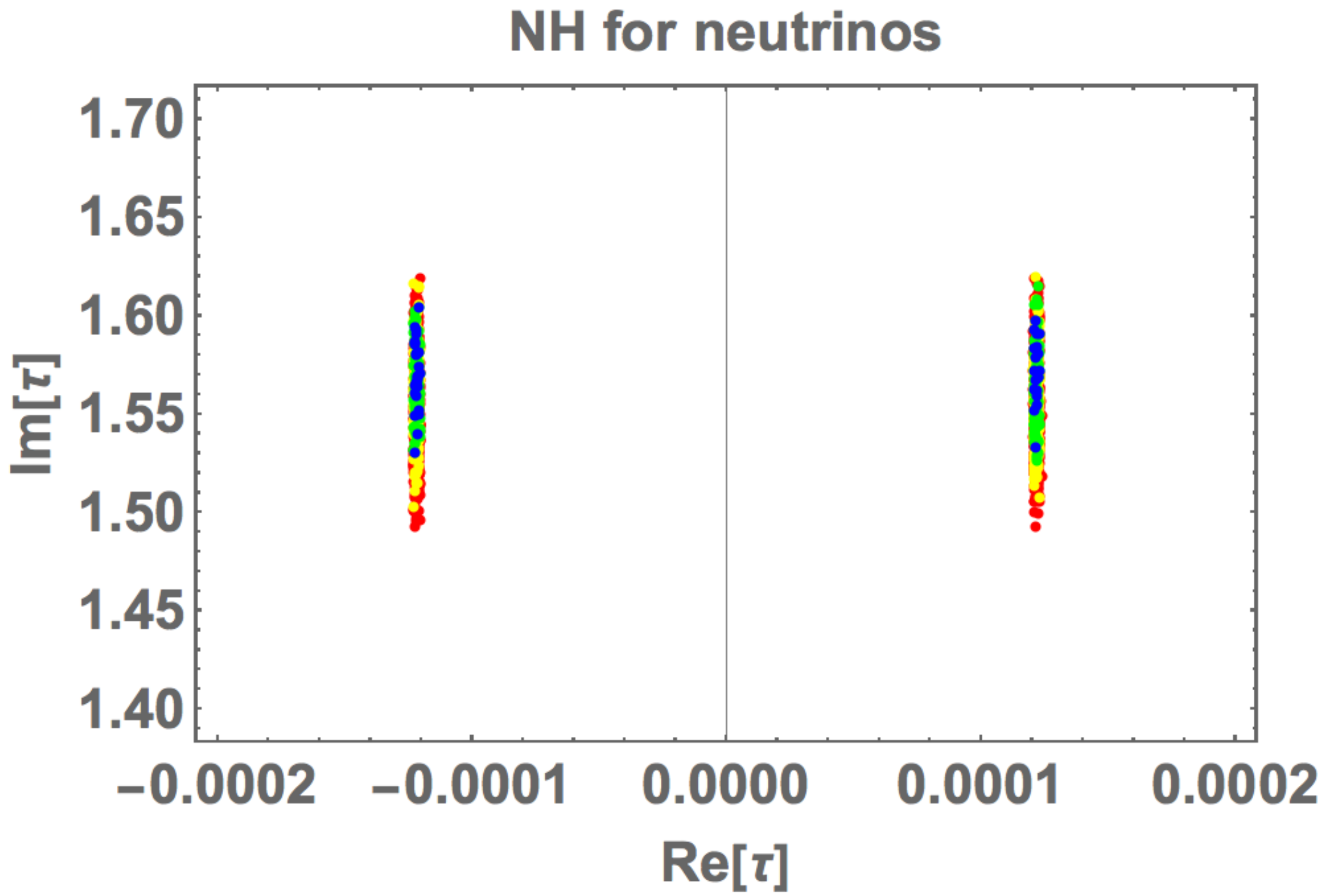}
\caption{Allowed region of $\tau$ in NH, where the color of points corresponds to the range of $\sqrt{\chi^2}$ value such that blue: $\sqrt{\chi^2} \leq 1$, green: $1< \sqrt{\chi^2}\le 2$, yellow: $2< \sqrt{\chi^2}\le 3$, and red: $3< \sqrt{\chi^2}\le 5$.}   
\label{fig:tau}\end{center}\end{figure}

 Fig.~\ref{fig:tau} shows the allowed region of $\tau$.
We find that $\tau\simeq1.49 i -1.62i$ within $\sqrt{\chi^2}=5$, which would be nearby at a fixed point of $\tau=i\times\infty$. Here, the color of points corresponds to the range of $\sqrt{\chi^2}$ value such that blue: $\sqrt{\chi^2} \leq 1$, green: $1< \sqrt{\chi^2}\le 2$, yellow: $2< \sqrt{\chi^2}\le 3$, and red: $3< \sqrt{\chi^2}\le 5$.

\begin{figure}[tb]\begin{center}
\includegraphics[width=75mm]{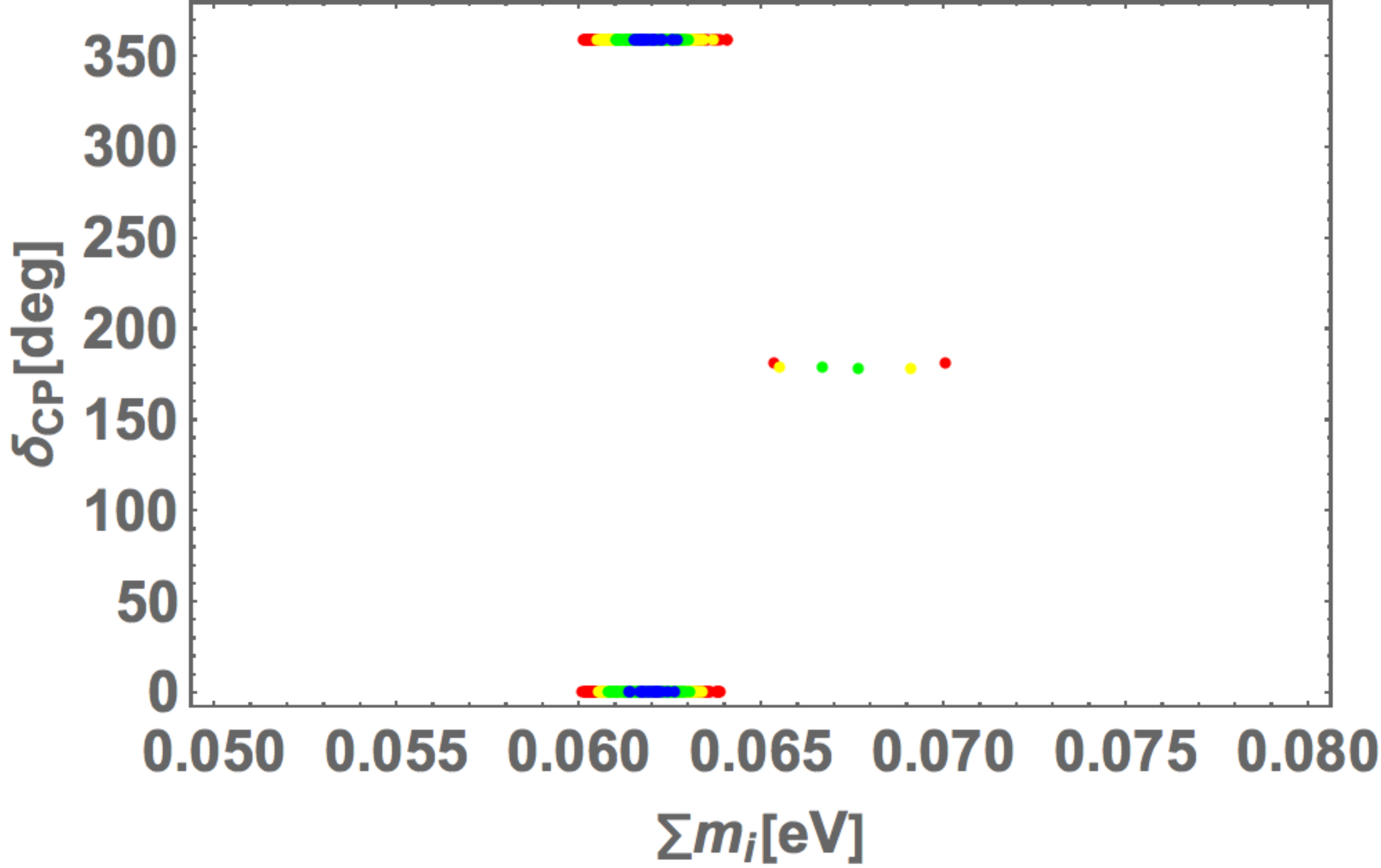}
\caption{Allowed region of $\sum m_i$ and Dirac CP phase $\delta_{\rm CP}$.
The color legend is the same as Fig.1.}   
\label{fig:phase-mass}\end{center}\end{figure}

Fig.~\ref{fig:phase-mass} shows the allowed region of $\sum m_i$ and Dirac CP phase $\delta_{\rm CP}$. 
We find that $\delta_{\rm CP}$ is localized at nearby $0$, then the allowed region of $\sum m_i$ is [60-64] meV within $\sqrt{\chi^2}=5$. On the other hand $\delta_{\rm CP}$ is localized at nearby $\pi$, the allowed region of $\sum m_i$ is [65-70] meV within $\sqrt{\chi^2}=5$.
Even though we do not show the figure on Majorana phases $\alpha_{31}$ and $\alpha_{21}$,
we obtained localized solutions at nearby $0$ or $\pi$ for  $\alpha_{31}$ and $\pi$ for  $\alpha_{21}$ only.
{Note that we have small CP violation in our allowed parameter region. This is due to small real part of $\tau$ which is only the source of CP violating phase.}

\begin{figure}[tb]\begin{center}
\includegraphics[width=75mm]{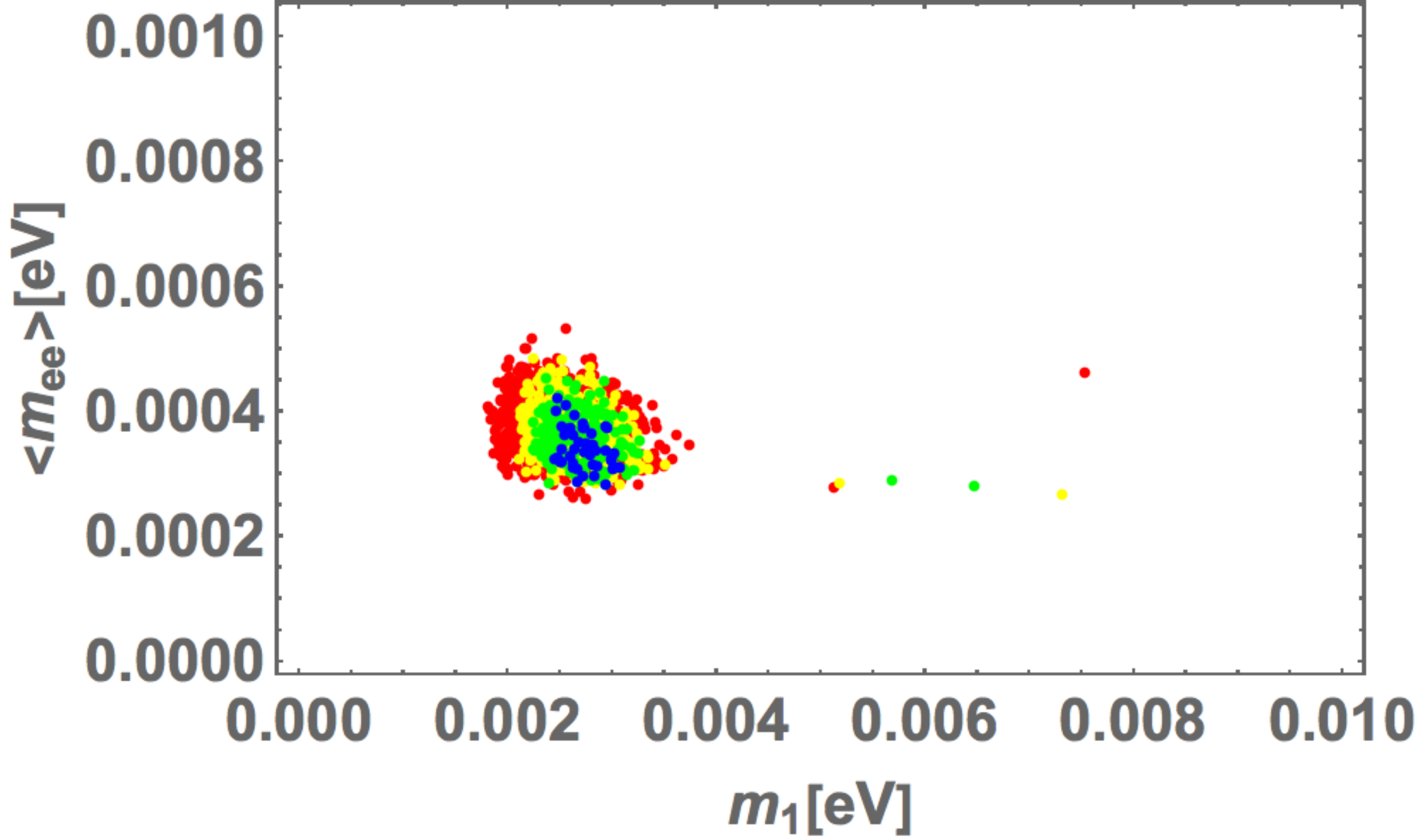}
\caption{Allowed region of the lightest neutrino mass $m_1$ and effective mass for neutrinoless double beta decay $\langle m_{ee}\rangle$.
The color legend is the same as Fig.1.}   
\label{fig:m1-mee}\end{center}\end{figure}

Fig.~\ref{fig:m1-mee} shows the allowed region of the lightest neutrino mass $m_1$ and effective mass for neutrinoless double beta decay $\langle m_{ee}\rangle$.
$m_1$ tends to be localized at the range of [2-4] meV within $\sqrt{\chi^2}=5$, while $\langle m_{ee}\rangle$ is localized at nearby [0.25-0.5] meV within $\sqrt{\chi^2}=5$.

\begin{table}[h]
	\centering
	\begin{tabular}{|c|c|c|} \hline 
			\rule[14pt]{0pt}{0pt}
 		&  NH  \\  \hline
			\rule[14pt]{0pt}{0pt}
		$\tau$ & $0.000122456 + 1.58071 i$       \\ \hline
		\rule[14pt]{0pt}{0pt}
%
		$[a_\eta, b_\eta,c_\eta]\times 10^6$ & $[95.067, 6.17325, 9.81507]$   \\ \hline
		\rule[14pt]{0pt}{0pt}
		$[m_N,B_{\tilde N}]/{\rm GeV}$ & $[8211.59, 302.961]$     \\ \hline
		\rule[14pt]{0pt}{0pt}
		$[m_{\psi_\chi},m_{\psi_\eta}]/{\rm GeV}$ & $[6075.09, 348.251]$     \\ \hline
		\rule[14pt]{0pt}{0pt}
		$[m_{H_1},m_{A_1},m_{H_2},m_{A_2}]/{\rm GeV}$ & $[7639.91, 384.573, 76.5929, 359.136]$    \\ \hline
		\rule[14pt]{0pt}{0pt}
		$\Delta m^2_{\rm atm}$  &  $2.52\times10^{-3} {\rm eV}^2$   \\ \hline
		\rule[14pt]{0pt}{0pt}
		$\Delta m^2_{\rm sol}$  &  $7.48\times10^{-5} {\rm eV}^2$        \\ \hline
		\rule[14pt]{0pt}{0pt}
		$\sin\theta_{12}$ & $ 0.550$   \\ \hline
		\rule[14pt]{0pt}{0pt}
		$\sin\theta_{23}$ &  $ 0.759$   \\ \hline
		\rule[14pt]{0pt}{0pt}
		$\sin\theta_{13}$ &  $ 0.149$   \\ \hline
		\rule[14pt]{0pt}{0pt}
		$[\delta_{CP}^\ell,\ \alpha_{21},\,\alpha_{31}]$ &  $[0.806^\circ,\, 180^\circ,\, 0.633^\circ]$   \\ \hline
		\rule[14pt]{0pt}{0pt}
		$\sum m_i$ &  $62.2$\,meV      \\ \hline
		\rule[14pt]{0pt}{0pt}
		$\langle m_{ee} \rangle$ &  $0.330$\,meV      \\ \hline
		\rule[14pt]{0pt}{0pt}
		$\sqrt{\Delta\chi^2}$ &  $0.536$     \\ \hline
		\hline
	\end{tabular}
	\caption{Numerical benchmark point of our input parameters and observables at nearby the fixed point $\tau= i \times \infty$ in NH. Here, the NH1 is taken such that $\sqrt{\Delta \chi^2}$ is minimum. On the other hand, NH2 is taken so that $\delta_{CP}$ is closest to the BF value of $195^\circ$ within $\sqrt{\Delta \chi^2}\le2$.}
	\label{bp-tab_nh}
\end{table}
%
Finally, we show a benchmark point for NH in Table~\ref{bp-tab_nh} that provide minimum $\sqrt{\chi^2}$ in our numerical analysis.

\section{Conclusion and discussion}
\label{sec:conclusion}

We have proposed a supersymmetric radiative seesaw model with modular $A_4$ symmetry.
Thanks to contributions of the SUSY partners, especially $\tilde N^c$, to the diagrams generating neutrino mass matrix, we have successfully constructed a predictive model in a minimum manner where we would not have obtained it in a non-SUSY model. 
Through our $\chi^2$ numerical analysis, we have obtained several allowed regions;
$\tau$ is localized at nearby $\tau\simeq1.49 i -1.62i$ within $\sqrt{\chi^2}=5$, which would be nearby at a fixed point of $\tau=i\times\infty$.
We find that $\delta_{\rm CP}$ is localized at nearby $0$ when the allowed region of $\sum m_i$ is [60-64] meV within $\sqrt{\chi^2}=5$
while $\delta_{\rm CP}$ is localized at nearby $\pi$ when the allowed region of $\sum m_i$ is [65-70] meV within $\sqrt{\chi^2}=5$.
Even though we do not show the figure on Majorana phases $\alpha_{31}$ and $\alpha_{21}$,
we have obtained localized solutions at nearby $0$ or $\pi$ for  $\alpha_{31}$ and $\pi$ for  $\alpha_{21}$ only.
$m_1$ tends to be localized at the range of [2-4] meV within $\sqrt{\chi^2}=5$, while $\langle m_{ee}\rangle$ is localized at nearby [0.25-0.5] meV within $\sqrt{\chi^2}=5$.

It would be worthwhile briefly mentioning the other aspects such as lepton flavor violations (LFVs) and dark matter(DM) candidate. 
Since input Yukawa couplings are typically of the order $10^{-6}$ and masses of particle mediating LFVs are of the order $100$ GeV at the lightest,
the typical branching ratio of the $\mu\to e\gamma$, which gives the most stringent constraint, is less than $10^{-20}$.
Therefore, it is totally safe for these kinds of constraints, because the upper bounds of $\mu\to e\gamma$ is $4.2\times 10^{-13}$.
{Notice here that we do not discuss LFVs induced by soft SUSY breaking terms in the MSSM assuming these couplings are small enough since it is beyond our scope.}

We have several DM candidates such as the lightest $N^c$, $\eta_{1,2}, \chi$, and their super-partners.
At first, $N^c$ might not be a good DM candidate since magnitude of associated Yukawa couplings are order of $10^{-6}$ at most, and the DM annihilation cross section   
to obtain the relic density would be too small. Thus, we may need to rely on $\eta_{1,2},\ \chi$ in obtaining observed relic density. Suppose $\eta_{1,2}$ does not mix with $\chi$,
$\eta_{1,2}$ cannot have mass difference between their neutral components in case of bosons.
Thus, $\eta_{1,2}$ would be ruled out as DM candidate from direct detection experiments since DM-nucleon scattering via running $Z$ boson is large.
Namely, pure $\chi$ would be a good DM candidate. The main annihilation modes to explain the observed relic density would arise from Higgs potential.
It would be easy to evade from constraints from direct detections via Higgs portals~\cite{Kanemura:2010sh}.  
Notice here that we do not discuss a DM analysis including super-partners that is beyond our scope and will be considered in future work.

\section*{Acknowledgments}
The work of H.O. is supported by the Junior Research Group (JRG) Program at the Asia-Pacific Center for Theoretical
Physics (APCTP) through the Science and Technology Promotion Fund and Lottery Fund of the Korean Government and was supported by the Korean Local Governments-Gyeongsangbuk-do Province and Pohang City.
The work is also supported by the Fundamental Research Funds for the Central Universities (T.~N.).
H.O. is sincerely grateful for all the KIAS members.

\appendix

\section{Formulas in modular $A_4$ framework}

Here we summarize some formulas of $A_4$ modular symmetry framework. 
Modular forms are 
holomorphic functions of modulus $\tau$, $f(\tau)$, which are transformed by
\begin{align}
& \tau \longrightarrow \gamma\tau= \frac{a\tau + b}{c \tau + d}\ ,~~ {\rm where}~~ a,b,c,d \in \mathbb{Z}~~ {\rm and }~~ ad-bc=1,  ~~ {\rm Im} [\tau]>0 ~, \\
& f(\gamma\tau)= (c\tau+d)^k f(\tau)~, ~~ \gamma \in \Gamma(N)~ ,
\end{align}
where $k$ is the so-called as the  modular weight.

A superfield $\phi^{(I)}$  is transformed under the modular transformation as 
\begin{equation}
\phi^{(I)} \to (c\tau+d)^{-k_I}\rho^{(I)}(\gamma)\phi^{(I)},
\end{equation}
where  $-k_I$ is the modular weight and $\rho^{(I)}(\gamma)$ represents an unitary representation matrix  corresponding to $A_4$ transformation.
Thus superpotential is invariant if sum of modular weight from fields and modular form in corresponding term is zero (also it should be invariant under $A_4$ and gauge symmetry).

The basis of modular forms is weight 2, $ Y^{(2)}_3=  (y_{1},y_{2},y_{3})$,  transforming
as a triplet of $A_4$ that is written in terms of the Dedekind eta-function $\eta(\tau)$ and its derivative \cite{Feruglio:2017spp}:
\begin{eqnarray} 
\label{eq:Y-A4}
y_{1}(\tau) &=& \frac{i}{2\pi}\left( \frac{\eta'(\tau/3)}{\eta(\tau/3)}  +\frac{\eta'((\tau +1)/3)}{\eta((\tau+1)/3)}  
+\frac{\eta'((\tau +2)/3)}{\eta((\tau+2)/3)} - \frac{27\eta'(3\tau)}{\eta(3\tau)}  \right), \nonumber \\
y_{2}(\tau) &=& \frac{-i}{\pi}\left( \frac{\eta'(\tau/3)}{\eta(\tau/3)}  +\omega^2\frac{\eta'((\tau +1)/3)}{\eta((\tau+1)/3)}  
+\omega \frac{\eta'((\tau +2)/3)}{\eta((\tau+2)/3)}  \right) , \label{eq:Yi} \\ 
y_{3}(\tau) &=& \frac{-i}{\pi}\left( \frac{\eta'(\tau/3)}{\eta(\tau/3)}  +\omega\frac{\eta'((\tau +1)/3)}{\eta((\tau+1)/3)}  
+\omega^2 \frac{\eta'((\tau +2)/3)}{\eta((\tau+2)/3)}  \right)\,, \nonumber \\
 \eta(\tau) &=& q^{1/24}\Pi_{n=1}^\infty (1-q^n), \quad q=e^{2\pi i \tau}, \quad \omega=e^{2\pi i /3}.
\nonumber
\end{eqnarray}
%
Modular forms with higher weight can be obtained from $y_{1,2,3}(\tau)$.
Some $A_4$ singlet modular forms used in our analysis are summarized as 
\begin{equation}
Y_1^{(4)} = y_1^2 + 2 y_1 y_3, \quad Y^{(6)}_{\bf 1}= y_1^2 + y_2^2 + y_3^2 - 3 y_1 y_2 y_3, \\
\end{equation}
where number in superscript indicates modular weight.

\end{document}